\documentclass[aps,prb,twocolumn,superscriptaddress,showpacs]{revtex4}
\usepackage{graphicx}
\usepackage{dcolumn}
\usepackage{bm}

\renewcommand{\v}[1]{{\bf #1}}
\newcommand{\nn}{\nonumber\\}
\newcommand{\be}{\begin{equation}}
\newcommand{\ee}{\end{equation}}
\newcommand{\ba}{\begin{eqnarray}}
\newcommand{\ea}{\end{eqnarray}}

\begin{document}

\title{Bond electronic polarization induced by spin}
\author{Chenglong Jia}
%\email[Electronic address:$~$]{cljia@skku.edu}
\affiliation{Department of Physics and Institute for Basic Science Research, \\
Sungkyunkwan University, Suwon 440-746, Korea}
\author{Shigeki Onoda}
\affiliation{Spin Superstructure Project, ERATO, Japan Science and
Technology Agency, c/o Department of Applied Physics, University of
Tokyo, Tokyo 113-8656, Japan}
%\email[Electronic address:$~$]{cljia@skku.edu}
\author{Naoto Nagaosa}
\affiliation{Spin Superstructure Project, ERATO, Japan Science and
Technology Agency, c/o Department of Applied Physics, University of
Tokyo, Tokyo 113-8656, Japan}
\affiliation{CREST, Department of Applied Physics, University of
Tokyo, Tokyo 113-8656, Japan} \affiliation{Correlated Electron
Research Center, National Institute of Advanced Industrial Science
and Technology, Tsukuba, Ibaraki 305-8562, Japan}
\author{Jung Hoon Han}
\email[To whom correspondence should be addressed. Electronic
address:$~$]{hanjh@skku.edu}
\affiliation{Department of Physics and Institute for Basic Science Research, \\
Sungkyunkwan University, Suwon 440-746, Korea} \affiliation{CSCMR,
Seoul National University, Seoul 151-747, Korea}
\date{\today}

\begin{abstract}
We study theoretically the electric polarization induced by
noncollinear spin configurations in the limit of strong
Hund coupling.
We employ a model of two magnetic ions sandwitching an oxygen ion.
It is shown that there appears a longitudinal polarization $P_x$
along the bond, which is roughly proportional to
$(m_r^x)^2-(m_l^x)^2$ where $m_{r(l)}^x$ is the
$x$-component of the spin orientation vector at right
(left) magnetic ion. A numerical study of the model Hamiltonian
yields both longitudinal and transverse electric dipole moments.
The transverse polarization is shown to have
a non-uniform as well as a uniform component,
with the latter being consistent with the previous theory.
The longitudinal polarization is non-uniform and oscillating
with the period half that of the spin order,
but the local magnitude is typically much larger than
the uniform transverse polarization
and may be detected by X-ray/neutron scattering experiments.
\end{abstract}

\maketitle

\section{Introduction}

A number of recent experimental breakthroughs has revived interests
in the phenomena of coupling of magnetic and electric (dipolar)
degrees of freedom in a class of materials known as
``multiferroics"\cite{NVO,TbMnO3-1,TbMnO3-2,hur,yamasaki,arima1,arima2}.
Some noteworthy observations include the development of dipole
moments accompanying the collinear-to-helical spin
ordering\cite{NVO,TbMnO3-1,arima2,yamasaki} and adiabatic control of
dipole moments through sweeping of applied magnetic fields
\cite{TbMnO3-2,hur}, which all unambiguously point to the strong
coupling of electric and magnetic degrees of freedom in these
compounds. A number of phenomenological\cite{mostovoy,harris} and
microscopic\cite{KNB,dagotto} theories has been advanced to
establish the connection between noncollinear spin order and
ferroelectricity.

In particular the work of Katsura, Nagaosa, and Balatsky
(KNB)\cite{KNB} proposed a microscopic theory
for the interplay between non-collinear magnetic order and
the dipolar polarization of the electronic wave function
induced by it. The magnetic (M) ion is modeled by three
degenerate $t_{2g}$ levels experiencing some external
magnetic field (to guarantee magnetic order) and subject to
spin-orbit coupling. Two such magnetic ions are bridged by
an intermediate oxygen (O) atom which itself has no
spin-orbit interaction. Solving the model Hamiltonian
perturbatively in the M-O hybridization amplitude, KNB
finds an electronic polarization orthogonal to the M-O-M
axis in the ground states of one and two holes.

In this paper we re-visit the M-O-M cluster model of KNB,
but in the different limit of a strong Hund coupling. The
M-O-M model in this limit is exactly solvable in the
absence of spin-orbit coupling, with two classes of
degenerate eigenstates. For these states there is no net
electric polarization. The spin-orbit coupling on the
magnetic sites is then introduced as a perturbation within
each degenerate manifold. The problem is exactly solvable
again and we can calculate the polarization $\langle \v r
\rangle$ for each of the eigenstates thus obtained. It is
shown that \textit{a non-zero polarization develops along
the direction of the M-O-M cluster}. We denote such
polarization as ``longitudinal", to distinguish it from the
``transverse" polarization, perpendicular to the cluster
axis, obtained in previous
theories\cite{KNB,dagotto,mostovoy}.

Numerical study of the mean-field Hamiltonian for the cluster
confirms the existence of longitudinal polarization predicted
analytically, and reveals a non-uniform component
in the transverse polarization which is unexpected in previous
theories\cite{KNB,dagotto,mostovoy}. Namely, oscillating
components appear in both the longitudinal and transverse
polarizations with a vanishing macroscopic average
whereas the polarization predicted by existing theories
gives a ``uniform'' component induced by a non-collinear spin order.

%%%%%%%%%%%%%%%%%%%%%%%%%%%%%%%%%%%%%%%%%%%%%%%%%%%%%%%%%%%%%%%%%%%%%%%%%%%%%%%%%%%%%%%%
\begin{figure}[h]
\begin{center}
\includegraphics[width=5cm]{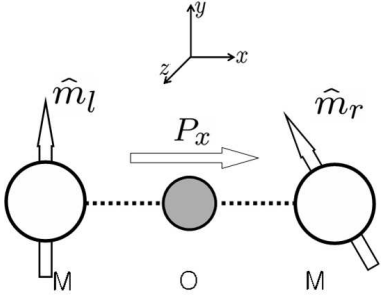}
\end{center}
\caption{Arrangement of two magnetic ions and an oxygen atom in the
M-O-M cluster. $\hat{m}_l$ and $\hat{m}_r$ are the magnetic moment
orientations of the magnetic sites bridged to the intervening oxygen
atom. Polarization $P_x$ along the cluster's axis  is found (see
text for details). } \label{model}
\end{figure}
%%%%%%%%%%%%%%%%%%%%%%%%%%%%%%%%%%%%%%%%%%%%%%%%%%%%%%%%%%%%%%%%%%%%%%%%%%%%%%%%%%%%%%%%%

This paper is organized as follows. In Sec.
\ref{eff-theory}, we introduce the mean-field Hamiltonian
for the three-site cluster, and in the limit of a large
Hund coupling, its approximate eigenstates are derived
analytically in a perturbation theory with respect to the
spin-orbit interaction. Electronic polarization is then
calculated for these eigenstates to the lowest order in the
spin-orbit coupling. In Sec. \ref{numerical-results},
numerical solution of the mean-field Hamiltonian is
presented. Fourier analysis of the polarization reveals the
existence of both uniform and non-uniform components in the
transverse part, and non-uniform component in the
longitudinal part. We conclude with a discussion in Sec.
\ref{discussion}. Some technical details of the derivations
used in Sec. \ref{eff-theory} are described in the
Appendix.
\\

\section{Model and Analytic results}
\label{eff-theory}

As a minimal unit giving rise to spin-polarization coupling
KNB introduced the model shown in Fig. \ref{model}. To
guarantee that a net average magnetic moment exists, each
magnetic site is subject to a spin polarization
field\cite{KNB}

\be H_{M} = -U \sum_{a=l,r}\hat{m}_a \cdot \left(
\sum_{b=xy,yz,zx}\v S_{a,b} \right) \label{H-M}\ee
where $a=l,r$ refers to left and right magnetic site in the
M-O-M cluster as depicted in Fig. \ref{model}. Spin-orbit
coupling is assumed to exist for the magnetic ions which
are connected to the oxygen through the wave function
overlap. The spins $\v S_{a,b}$ in each of the three
degenerate $t_{2g}$ orbitals are subject to the local
magnetic field along $\hat{m}_a = (\sin\theta_a \cos \phi_a
, \sin\theta_a \sin\phi_a , \cos \theta_a)$. Ultimately
such effective fields arise due to inter-atomic exchange
interaction but also due to the strong Hund coupling within
the individual magnetic ion.  In such a case the strength
of $U$ mimicking the Hund coupling energy may well exceed
the spin-orbit interaction strength $\lambda$. The
large-$U$ limit thus provides a natural starting point for
the analysis of the spin-polarization coupling in the
cluster model. In this section we propose a method that is
particularly suited to treat the large $U$ limit, namely $U
\gg \lambda, V$ where $V$ represents the M-O hybridization.

The overall Hamiltonian describing the three-atom cluster
is given by $H = H_{M}+H_{O} + H_{V}$:  $H_M$ is already
given in Eq. (\ref{H-M}) and the other two terms, referring
to the oxygen orbital and the M-O hybridization, are

\ba   && H_{O} = E_{p} \sum_{b=x,y,z} \sum_{\sigma }
p_{b\sigma }^{+}p_{b\sigma }  \nn
&& H_{V} = V\sum_{\sigma} (d_{l,xy\sigma}^{+}p_{y\sigma}+
d_{l,zx\sigma}^{+}p_{z\sigma} \nn && ~~~~~~~~~~~~
-d_{r,xy\sigma}^{+}p_{y\sigma} -
d_{r,zx\sigma}^{+}p_{z\sigma}+ \mathrm{h.c.} ).  \nn
\label{MOM-full-H}\ea
Particle-hole transformations have been implemented on all
three atomic sites\cite{KNB}. The on-site energy of the
oxygen atom, $-E_p$, becomes $+E_p$ after the
transformation. The spin-orbit interaction, not included
here, will be introduced later as a perturbation.

Now we discuss how one obtains a low-energy effective
Hamiltonian from Eq. (\ref{MOM-full-H}). First, due to the
large energy difference $2U$ separating the spin-up
(parallel to $\hat{m}_a$) and spin-down (antiparallel to
$\hat{m}_a$) $t_{2g}$ states, we can truncate the
high-energy $d$-orbital states from the outset and write
down the effective Hamiltonian using only the low-energy
operators

\ba d_{a,xy}^{+} &=&\cos {\frac{\theta
_{a}}{2}}d_{a,xy\uparrow }^{+}+e^{-i\phi _{a}}\sin
{\frac{\theta _{a}}{2}}d_{a,xy\downarrow }^{+}  \nn
d_{a,zx}^{+} &=&\cos {\frac{\theta _{a}}{2}}d_{a,zx\uparrow
}^{+}+e^{-i\phi _{a}}\sin {\frac{\theta
_{a}}{2}}d_{a,zx\downarrow }^{+} . \label{rotated-spins}
\ea
The $d_{yz}$ orbital does not hybridize with any of the
oxygen $p$-orbitals in the linear geometry we consider in
this paper. This, and the $p_x$-orbital which does not
hybridize with any of the $d$-orbitals, will be left out.

In this reduced Hilbert space the low-energy effective
Hamiltonian becomes

\ba   && H' ~ = ~ H'_{M}+H'_{O} + H'_{V}  \nn
&& H'_{M} = -U \sum\limits_{a=l,r}\left(
d_{a,xy}^{+}d_{a,xy}+d_{a,zx}^{+}d_{a,zx} \right) \nn &&
H'_{O} = E_{p} \sum_{\sigma } \left( p_{y\sigma
}^{+}p_{y\sigma }+p_{z\sigma }^{+}p_{z\sigma } \right)  \nn
&& H'_{V} =V( d_{l,xy}^{+}p_{l,y}+d_{l,zx}^{+}p_{l,z} \nn
&& ~~~~~~~~~- d_{r,xy}^{+}p_{r,y}-d_{r,zx}^{+}p_{r,z}
+\mathrm{h.c.}) . \nn \label{MOM-H}\ea
In writing down $H'_V$ spin rotations similar to Eq.
(\ref{rotated-spins}) have been employed also for the
$p$-orbital operators:

\ba p_{a,y}^{+} &=&\cos {\frac{\theta _{a}}{2}}p_{y\uparrow
}^{+}+e^{-i\phi _{a}}\sin {\frac{\theta
_{a}}{2}}p_{y\downarrow }^{+}  \nn p_{a,z}^{+} &=&\cos
{\frac{\theta _{a}}{2}}p_{z\uparrow }^{+}+e^{-i\phi
_{a}}\sin {\frac{\theta _{a}}{2}}p_{z\downarrow }^{+} . \ea
The axis of the three-atom cluster is taken as the
$+\hat{x}$ axis going from left to right in Fig.
\ref{model}. Until the spin-orbit coupling is introduced,
the $(d_{xy},p_y)$ orbital pair remains decoupled from
$(d_{zx},p_z )$.

The $p$-orbital operators $(p_{l,y}, p_{r,y})$ are not orthogonal to each
other. The same is true of $(p_{l,z}, p_{r,z})$. For further manipulation of
the Hamiltonian we need to introduce a set of orthogonal operators

\ba Y_{1}^{+} &=&\frac{1}{\sqrt{2(1+|\kappa
|)}}(p_{r,y}^{+}+e^{i\eta
}p_{l,y}^{+})  \nonumber \\
Y_{2}^{+}&=&\frac{1}{\sqrt{2(1-|\kappa |)}} (p_{r,y}^{+}-e^{i\eta
}p_{l,y}^{+})  \nonumber \\
Z_{1}^{+} &=&\frac{1}{\sqrt{2(1+|\kappa |)}}(p_{r,z}^{+}+e^{i\eta
}p_{l,z}^{+})  \nonumber \\
Z_{2}^{+}&=&\frac{1}{\sqrt{2(1-|\kappa |)}}
(p_{r,z}^{+}-e^{i\eta }p_{l,z}^{+}) \ea
for the $p$-orbitals and

\ba
D_{1,xy}^{+}&=&{\frac{1}{\sqrt{2}}}(-d_{r,xy}^{+}+e^{i\eta
}d_{l,xy}^{+}) \nn
D_{2,xy}^{+}&=&{\frac{1}{\sqrt{2}}}
(-d_{r,xy}^{+}-e^{i\eta}d_{l,xy}^{+}) \nn
D_{1,zx}^{+}&=&{\frac{1}{\sqrt{2}}}(-d_{r,zx}^{+}+e^{i\eta}
d_{l,zx}^{+}) \nn
D_{2,zx}^{+}&=&{\frac{1}{\sqrt{2}}}
(-d_{r,zx}^{+}-e^{i\eta }d_{l,zx}^{+}) \ea
for $d$-orbital states. We have defined $e^{i\eta }=\kappa
/|\kappa |$ where

\ba \kappa &=&\langle p_{l,y}|p_{r,y}\rangle =\langle
p_{l,z}|p_{r,z}\rangle \nn
&=&\cos {\frac{\theta_{l}}{2}}\cos
{\frac{\theta_{r}}{2}}+e^{i(\phi _{l}-\phi _{r})}\sin
{\frac{\theta_{l}}{2}}\sin {\frac{\theta _{r}}{2}}. \ea
Using these new operators, the Hamiltonian in Eq.
(\ref{MOM-H}) becomes

\ba H &=&H_{Y}+H_{Z}  \nn
H_{Y} &=&-\sum_{\alpha =1,2}\left(
\begin{array}{cc}
D_{\alpha ,xy}^{+} & Y_{\alpha }^{+}\end{array} \right)
\mathcal{H}_\alpha \left(
\begin{array}{c}
D_{\alpha ,xy} \\
Y_{\alpha}\end{array} \right)  \nn
H_{Z} &=& -\sum_{\alpha=1,2}\left(
\begin{array}{cc}
D_{\alpha ,zx}^{+} & Z_{\alpha }^{+}\end{array} \right)
\mathcal{H}_\alpha \left(
\begin{array}{c}
D_{\alpha ,zx} \\
Z_{\alpha }
\end{array}
\right)  \nn
\mathcal{H}_\alpha & = &E_0 + E_\alpha \left(
\begin{array}{cc}
\cos \beta_{\alpha} & -\sin \beta_{\alpha } \\
-\sin \beta_{\alpha} & -\cos \beta_{\alpha }%
\end{array}\right) \ea
with $E_{0}=(U-E_{p})/2$, $E_{\alpha }=\sqrt{(U+E_{p})^2 /4
+V_{\alpha }^{2}} $, and $V_{1,2}=V\sqrt{1\pm |\kappa |}$,
$(\cos \beta _{\alpha }, \sin \beta_{\alpha })=
((U+E_{p})/2E_{\alpha }, V_{\alpha }/E_{\alpha })$. $H_{Y}$
and $H_{Z}$ can be diagonalized through the rotation

\ba
\left(
\begin{array}{c}
D_{\alpha ,xy} \\
Y_{\alpha } \end{array} \right) &=&\left(
\begin{array}{cc}
-\cos \beta _{\alpha }/ 2 & \sin \beta _{\alpha }/ 2 \\
\sin \beta _{\alpha }/ 2 & \cos \beta _{\alpha }/ 2%
\end{array}\right) \left(
\begin{array}{c}
\psi _{\alpha } \\
\varphi _{\alpha }\end{array} \right)  \nn
\left( \begin{array}{c}
D_{\alpha ,zx} \\
Z_{\alpha }
\end{array}
\right) &=&\left(
\begin{array}{cc}
-\cos \beta _{\alpha }/ 2 & \sin \beta _{\alpha }/ 2 \\
\sin \beta _{\alpha }/ 2 & \cos \beta _{\alpha }/ 2
\end{array}
\right) \left(
\begin{array}{c}
\psi _{\alpha }^{\prime } \\
\varphi _{\alpha }^{\prime }%
\end{array}\right) \nn\ea
into

\ba &&H_{Y} =\sum_{\alpha }-(E_{0}\!+\!E_{\alpha })\psi
_{\alpha }^{+}\psi _{\alpha
}+\sum_{\alpha}(E_{\alpha}\!-\!E_{0})\varphi_{\alpha
}^{+}\varphi _{\alpha }, \nn
&&H_{Z}
=\sum_{\alpha}-(E_{0}\!+\!E_{\alpha})\psi_{\alpha}^{\prime
+}\psi_{\alpha }^{\prime}+\sum_{\alpha}(E_{\alpha
}\!-\!E_{0})\varphi_{\alpha }^{\prime +}\varphi _{\alpha
}^{\prime }. \nn
\ea
Four degenerate levels $(\psi_1, \psi_2, \psi^{\prime }_1,
\psi^{\prime }_2)$ are separated from the other four
degenerate set $(\varphi_1, \varphi_2, \varphi^{\prime }_1,
\varphi^{\prime }_2)$ with an energy spacing of nearly
$U+E_p$. This large energy separation sets the stage for
introducing spin-orbit interaction $H_{SO} = \lambda S\cdot
L$ within each of the four-dimensional manifolds, but not
between the two manifolds. We obtain the effective
Hamiltonian valid within each manifold,

\ba && \mathcal{H} =\left(
\begin{array}{c}
\psi _{1}^{+} \\
\psi _{2}^{+} \\
\psi _{1}^{\prime +} \\
\psi _{2}^{\prime +}%
\end{array}
\right)^T \left(
\begin{array}{cccc}
-U_{1} & 0 & -i\lambda _{1} & -i\lambda _{2} \\
0 & -U_{2} & -i\lambda _{2} & -i\lambda _{3} \\
i\lambda _{1} & i\lambda _{2} & -U_{1} & 0 \\
i\lambda _{2} & i\lambda _{3} & 0 & -U_{2}%
\end{array}%
\right) \left(
\begin{array}{c}
\psi _{1} \\
\psi _{2} \\
\psi _{1}^{\prime } \\
\psi _{2}^{\prime }%
\end{array}%
\right)  \nn &&~~~+\left(
\begin{array}{c}
\varphi _{1}^{+} \\
\varphi _{2}^{+} \\
\varphi _{1}^{\prime +} \\
\varphi _{2}^{\prime +}%
\end{array}%
\right)^T \left(
\begin{array}{cccc}
E_{1}^{p} & 0 & -i\lambda _{1}^{\prime } & -i\lambda _{2}^{\prime } \\
0 & E_{2}^{p} & -i\lambda _{2}^{\prime } & -i\lambda _{3}^{\prime } \\
i\lambda _{1}^{\prime } & i\lambda _{2}^{\prime } & E_{1}^{p} & 0 \\
i\lambda _{2}^{\prime } & i\lambda _{3}^{\prime } & 0 & E_{2}^{p}%
\end{array}%
\right) \left(
\begin{array}{c}
\varphi _{1} \\
\varphi _{2} \\
\varphi _{1}^{\prime } \\
\varphi _{2}^{\prime }%
\end{array}%
\right)  \nn \ea
where $U_{\alpha }=E_{0}+E_{\alpha }$,
$E_{\alpha}^{p}=E_{\alpha }-E_{0}$, and

\ba i\lambda _{1} &=&\left\langle \psi _{1}^{\prime }|H_{SO}|\psi
_{1}\right\rangle =\frac{i\lambda }{2}\cos ^{2}\frac{\beta
_{1}}{2}\left( m_{r}^{x}+m_{l}^{x}\right)  \nn
i\lambda _{2} &=&\left\langle \psi _{1}^{\prime }|H_{SO}|\psi
_{2}\right\rangle = \left\langle \psi _{2}|H_{SO}|\psi _{1}^{\prime
}\right\rangle  \nn
&& =\frac{i\lambda }{2}\cos \frac{\beta _{1}}{2}\cos \frac{\beta
_{2}}{2} \left( m_{r}^{x}-m_{l}^{x}\right)  \nn
i\lambda _{3}&=& \left\langle \psi _{2}^{\prime }|H_{SO}|\psi
_{2}\right\rangle =\frac{i\lambda }{2}\cos ^{2}\frac{\beta
_{2}}{2}\left( m_{r}^{x}+m_{l}^{x}\right)  \nn
i\lambda _{1}^{\prime} &=& \left\langle \varphi _{1}^{\prime
}|H_{SO}|\varphi _{1}\right\rangle =\frac{i\lambda }{2}\sin
^{2}\frac{ \beta _{1}}{2}\left( m_{r}^{x}+m_{l}^{x}\right)  \nn
i\lambda_{2}^{\prime } &=& \left\langle \varphi _{1}^{\prime
}|H_{SO}|\varphi _{2}\right\rangle =\left\langle \varphi
_{2}|H_{SO}|\varphi _{1}^{\prime }\right\rangle  \nn &&
=\frac{i\lambda }{2}\sin \frac{\beta _{1}}{2}\sin \frac{\beta
_{2}}{2} \left( m_{r}^{x}-m_{l}^{x}\right)  \nn
i\lambda_{3}^{\prime } &=&\left\langle \varphi _{2}^{\prime
}|H_{SO}|\varphi _{2}\right\rangle =\frac{i\lambda }{2}\sin
^{2}\frac{\beta _{2}}{2}\left( m_{r}^{x}+m_{l}^{x}\right).
\nn \label{SO-elements}\ea
Here $m_{a}^{x}$ refers to the $x$-component of the local
quantization axis for left and right magnetic sites.
Because of small $V/U$, we can safely take
\begin{eqnarray}
E_{1}&=&E_{2}=E=(U+E_p)/2
\nonumber\\
\cos \beta_{\alpha }/2 &\approx&1
\nonumber\\
\sin \beta _{\alpha }/2 &\approx& V_{\alpha}/(U+E_{p})
\label{approx}
\end{eqnarray}
in Eq. (\ref{SO-elements}).  The above
Hamiltonian $\mathcal{H}$ can be diagonalized to the form
\be \mathcal{H}=\sum_{i=1}^4 E_{i}^{L}\cdot \Psi
_{i}^{+}\Psi _{i}+\sum_{i=1}^4 E_{i}^{H}\cdot \Phi
_{i}^{+}\Phi _{i} \ee
with the eigen-states

\begin{eqnarray}
\Psi _{i} &=&\xi _{1}^{i}\psi _{1}+\xi _{2}^{i}\psi _{2}+\xi _{3}^{i}\psi
_{1}^{\prime }+\xi _{4}^{i}\psi _{2}^{\prime }  \nonumber \\
\Phi _{i} &=&\zeta _{1}^{i}\varphi _{1}+\zeta _{2}^{i}\varphi _{2}+\zeta
_{3}^{i}\varphi _{1}^{\prime }+\zeta _{4}^{i}\varphi _{2}^{\prime }.
\end{eqnarray}
One can verify that the coefficients satisfy

\ba
&& |\xi _{1}^{i}|=|\xi _{3}^{i}|,~~|\xi _{2}^{i}|=|\xi _{4}^{i}|, ~~~%
\overline{\xi}_{1}^{i}\xi _{2}^{i}=\overline{\xi}_{3}^{i}\xi
_{4}^{i}\in \mathrm{Re}  \nn
&& |\zeta _{1}^{i}|=|\zeta _{3}^{i}|,~~|\zeta _{2}^{i}|=|\zeta _{4}^{i}|,~~~%
\overline{\zeta}_{1}^{i}\zeta _{2}^{i}=\overline{\zeta}_{3}^{i}\zeta
_{4}^{i}\in \mathrm{Re}. \ea
This completes the formal derivation of the full set of eigenstates in the
limit $U \gg \lambda, V$. The remaining task is to calculate the
polarization $\langle \v r \rangle$ for each of the eigenstates
obtained.

\begin{figure}[h]
\begin{center}
\includegraphics[width=5cm]{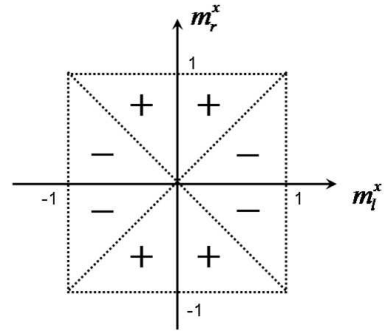}
\end{center}
\caption{The direction of longitudinal polarization $P_x = \v P\cdot
\hat{x} $ in the ground state wave function plotted for $(m^x_l ,
m^x_r)$, $-1 \le m^x_l  , m^x_r  \le 1$. +(-) refers to $P_x > (<)$
0. $P_x $ is indeterminate along two dotted lines ($m_r^x = m_l^x$
and $m_r^x = -m_l^x$) due to the level crossing occurring along
these lines. } \label{phase}
\end{figure}

Due to the shapes of $d$- and $p$-orbitals,
only the following overlap integrals are non-zero ($\v x = x \hat{x}$):

\ba &&\langle d_{r,xy\sigma }|\v x|p_{y\sigma }\rangle =\langle
d_{l,xy\sigma }|\v x|p_{y\sigma }\rangle   \nonumber \\
&=&\langle d_{r,zx\sigma }|\v x|p_{z\sigma }\rangle =\langle
d_{l,zx\sigma }|\v x|p_{z\sigma }\rangle =L\hat{x}.
\label{non-zero-overlap} \ea
Here $L=\int d^{3}\v r d_{i,xy\sigma }(\v r)\v x p_{y\sigma
}(\v r)$ ($i=r,l$). If non-zero polarization develops, it
can only be in the $\hat{x}$ direction, i.e. along the axis
of the three-atom cluster, within the lowest order in
$\lambda/\Delta$, $\Delta = U + E_p$. Note that the
transverse component obtained in Ref. \onlinecite{KNB}
corresponds to the terms of the first order in
$\lambda/\Delta$. This contribution is numerically studied
later in Sec.~\ref{numerical-results}. Analytic
calculations of the diagonal and off-diagonal contributions
to the polarization are given in Appendices. The final
expression for the polarization $\v P$ reads (details are
given in the Appendix)

\ba
\v {P}_{\Psi _{i}}&=&\langle \Psi _{i}|\v r|\Psi _{i}\rangle =2\bar{%
\xi}_{1}^{i}\xi _{2}^{i}\left[ \langle \psi _{1}|\v r|\psi
_{2}\rangle +\langle \psi _{1}^{\prime }|\v r|\psi _{2}^{\prime
}\rangle \right] \nn &\approx &\hat{x}\times
\frac{8\bar{\xi}_{1}^{i}\xi _{2}^{i}LV}{\Delta }, \nn \nn
\v {P}_{\Phi _{i}}&=&\langle \Phi _{i}|\v r|\Phi _{i}\rangle =\frac{1%
}{2}\bar{\zeta}_{1}^{i}\zeta _{2}^{i}\left[ \langle \varphi _{1}|\v
r|\varphi _{2}\rangle +\langle \varphi  _{1}^{\prime }|\v r|\varphi
_{2}^{\prime }\rangle \right]   \nn &\approx& -\hat{x}\times
\frac{\sqrt{2}\bar{\zeta}_{1}^{i}\zeta _{2}^{i}LV}{\Delta
}\sqrt{1-\sigma _{1}\cdot \sigma _{2}}. \ea
Note that $\bar{\xi}_{1}^{i}\xi _{2}^{i}$ and
$\bar{\zeta}_{1}^{i}\zeta _{2}^{i}$ are also dependent on the spin
configuration. The full spin dependence of the polarization
may be rather complicated. However, because of small $V/U$, we can
take $U_{1}=U_{2}=U$ and

\ba \lambda _{1} &=& \lambda_3 = \frac{\lambda }{2}( m^x_l + m^x_r )
\nn \lambda_{2} &=& \frac{\lambda }{2}( m^x_r - m^x_l ).
\label{lambda-formula} \ea
Under such an approximation the low-energy eigenstates are given by
the surprisingly simple form

\ba   && \Psi _{1}=(i\psi _{1}+i\psi _{2}+\psi _{1}^{\prime
}+\psi _{2}^{\prime })/2, ~~~ E_{1}^{L} = -U
\!-\!\lambda_{1}\!-\!\lambda _{2}  \nn
&& \Psi _{2}=(i\psi _{1}-i\psi_{2}-\psi _{1}^{\prime }+\psi
_{2}^{\prime })/2,~~~ E_{2}^{L
}=-U\!+\!\lambda_{1}\!-\!\lambda_{2}  \nn
&& \Psi_{3}=(-i\psi _{1}+i\psi _{2}-\psi
_{1}^{\prime}+\psi_{2}^{\prime })/2, ~
E_{3}^{L}=-U\!-\!\lambda_{1}\!+\!\lambda_{2}  \nn
&& \Psi_{4}=(-i\psi _{1}-i\psi _{2}+\psi _{1}^{\prime
}+\psi_{2}^{\prime })/2, ~
E_{4}^{L}=-U\!+\!\lambda_{1}\!+\!\lambda_{2}.  \nn
\label{eigensystem}\ea
For each of the eigenstates above the polarization $\v P_{\Psi_{i}}$
reads

\ba
\v {P}_{\Psi _{1}} &=&\v {P}_{\Psi _{4}}=\hat{x}\times \frac{2LV}{%
\Delta }  \nn
\v {P}_{\Psi _{2}} &=&\v {P}_{\Psi _{3}}=-\hat{x}\times \frac{2LV}{%
\Delta }.  \label{P} \ea
The polarization, constant in magnitude but reversible in
sign, is developed along the M-O-M cluster. Since $\lambda
_{1}$ and $\lambda _{2}$ depend on the spin directions and
in particular on $m^x_l$ and $m^x_r$ as shown in Eq.
(\ref{lambda-formula}), which one of the four states shown
in Eq. (\ref{eigensystem}) becomes the true ground state
depends on the set of values $(m_{l}^{x},m_{r}^{x})$. In
Fig. \ref{phase} we map out the polarization directions of
the ground state for all the available situations
within the zeroth order in $\lambda/\Delta$. The
reversal of the polarization direction occurs whenever
$|m_{l}^{x}|=|m_{r}^{x}|$ line is crossed. The sudden
change of the sign is due to level crossing of the two
eigenstates when $\lambda_1$ or $\lambda_2$ vanishes at
$|m^x_l | = |m^x_r |$. The ground state polarization is
consistent with the functional form $P_x \sim \left( m_r^x
\right)^2 - \left( m_l^x \right)^2 $.

If a single hole is introduced in the cluster, the
polarization is given by one of the expressions in Eq.
(\ref{P}). With two holes, two opposite polarizations from
Eq. (\ref{P}) cancel out to produce $\v P = 0$. In the
numerical results presented in the next section, however,
one finds a finite longitudinal polarization even for the
two-hole case because of the higher-order terms in $\lambda/\Delta$,
and the polarization direction is still
dictated by $P_x \sim \left( m_r^x \right)^2 - \left( m_l^x
\right)^2 $.

\section{Numerical Results}
\label{numerical-results}

The results of the previous section gave the electronic polarization
solely along the axis of the atomic cluster. In the analysis it was
assumed that the large effective Zeeman energy $U$ dominates over
both the spin-orbit interaction $\lambda$ and the hybridization
amplitude $V$. It was also assumed that $d_{yz}$ and $p_x$ orbitals
did not play a role. According to Ref. \onlinecite{KNB}, on the
other hand, transverse polarization such as $P_y$ can only arise
from the overlap integral $\langle d_{yz} | y | p_z \rangle$ or
$\langle d_{xy} | y | p_x \rangle$. In the numerical approach this
condition can be relaxed and indeed we find non-zero transverse
polarization as well as the longitudinal polarization in accordance
with the theory presented in the previous section. In general the
results depend sensitively on the number of holes inserted in the
cluster, with the even and odd number of holes yielding different
behavior. As a representative case for each, we consider one- and
two-hole situations in detail. When extrapolated to the infinite
lattice they correspond to the metallic and insulating cases,
respectively.

The Hilbert space for the single-particle mean-field Hamiltonian is
18-dimensional, consisting
of Eq. (\ref{H-M}) and Eq. (\ref{MOM-full-H}), with six from each
atomic site. In practice, one can treat the 16-dimensional problem
leaving out the oxygen  $p_x$ orbital without loss of generality,
since it remains decoupled from the rest of the atomic states in the
exact solution.

\subsection{One hole (double-exchange interaction)}
\begin{figure}[t]
\begin{center}
\includegraphics[width=7cm]{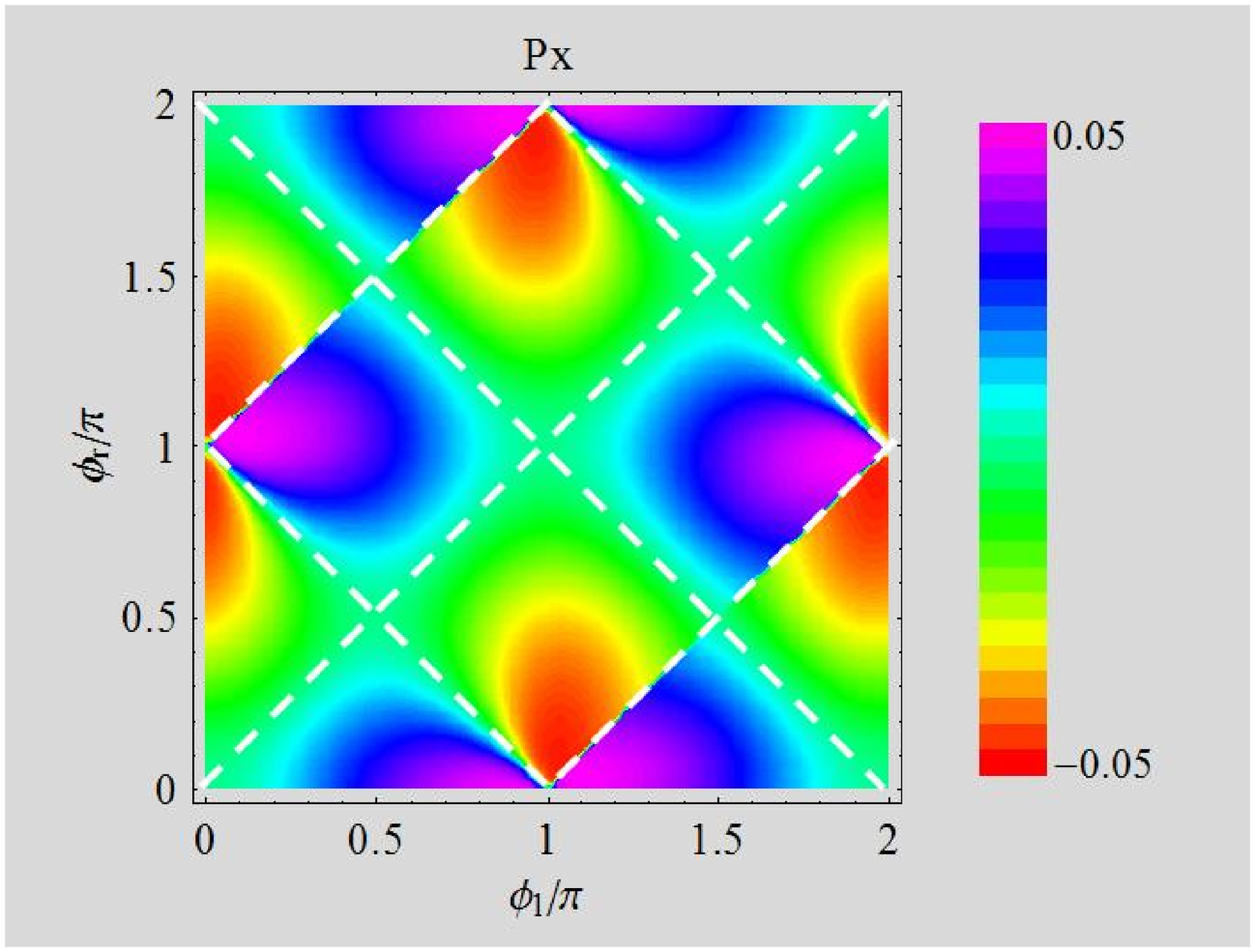}
\includegraphics[width=7cm]{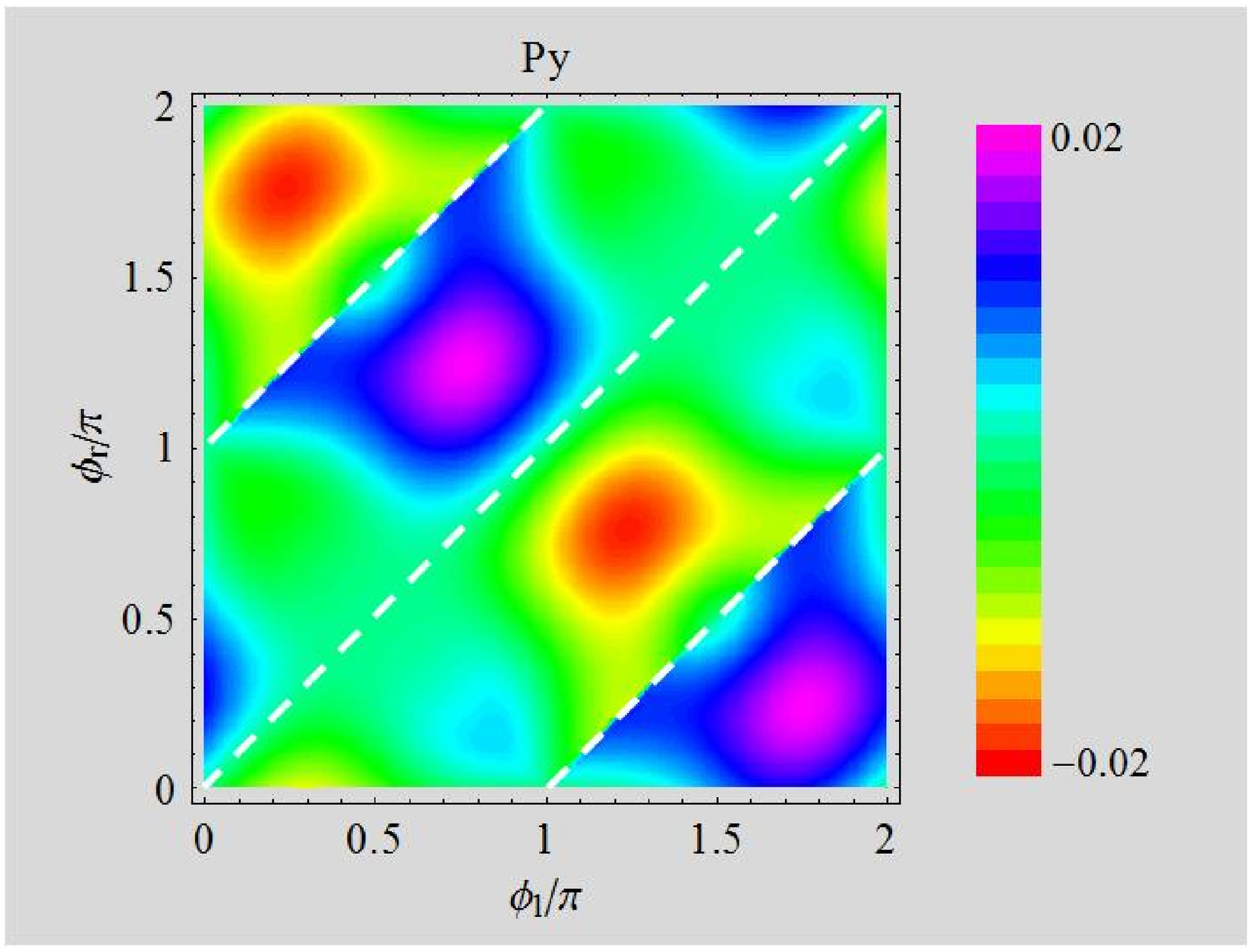}
\end{center}
\caption{ (color online) The calculated polarization $P_x$ and $P_y$
from exact diagonalization of the full M-O-M cluster Hamiltonian
with a single hole. Dotted white lines indicate $P_x$ or $P_y$ equal
to zero. $P_x$ vanishes not only when the spins are collinear, but
also when $\phi_l + \phi_r = \pm \pi$, as discussed in
Sec.~\ref{eff-theory}.
 The overall directional dependence
for $P_x$ is consistent with $\cos 2\phi_r - \cos 2\phi_l$. The
sharp change in the sign of $P_x$ along $\phi_r =\phi_l \pm \pi$
lines is due to the energy level crossing in the ground state. $P_x$
turns to zero along the lines (indicated as dotted white lines)
where the spins become collinear. The parameters used in this
calculation are $(U/V, E_p /V ,\lambda /V) = (10,30,0.3)$. }
\label{polarization}
\end{figure}

The polarization $\langle \v r \rangle$ is calculated with respect
to the lowest-energy eigenstate of the particle-hole transformed
Hamiltonian corresponding to putting one hole in the cluster. For
ease of presentation we restrict ourselves to situations where both
spins lie in the XY plane: $\hat{m}_l = ( \cos \phi_l, \sin\phi_l,
0)$, $\hat{m}_r = ( \cos \phi_r, \sin\phi_r, 0)$. The resulting
longitudinal polarization $P_x$ is plotted as a function of $(\phi_l
, \phi_r )$ in Fig. \ref{polarization}. The similarity between the
numerically obtained $P_x$ and the predicted behavior $P_x \sim \cos
2\phi_r - \cos 2\phi_l$ is striking. The magnitude of the
polarization, according to Eq. (\ref{P}) and using the parameters
used in the numerical study, is expected to be $|P_x | \sim 0.05 L$,
in good agreement with the numerical results.

In addition, we find non-zero $P_y$ in the numerical calculation due
to the inclusion of $d_{yz}$ orbital. The behavior of $P_y$ we find
is consistent with the directional dependence $P_y \sim
\mathrm{sgn}[\cos(\Delta\phi/2)]\sin(\Delta\phi/2)$ $(\Delta\phi =
\phi_r - \phi_l)$ expected in Ref. \onlinecite{KNB} (See Eq. (4) in
Ref. \onlinecite{KNB}). Also the order of magnitude $|P_y | \sim
0.01 L$ obtained numerically is consistent with Ref. \onlinecite{KNB}.
$P_z$ vanishes for the two spins lying in the $XY$ plane, as shown
both analytically and numerically.

\subsection{Two holes (superexchange interaction)}

\begin{figure}[t!]
\begin{center}
\includegraphics[width=7cm]{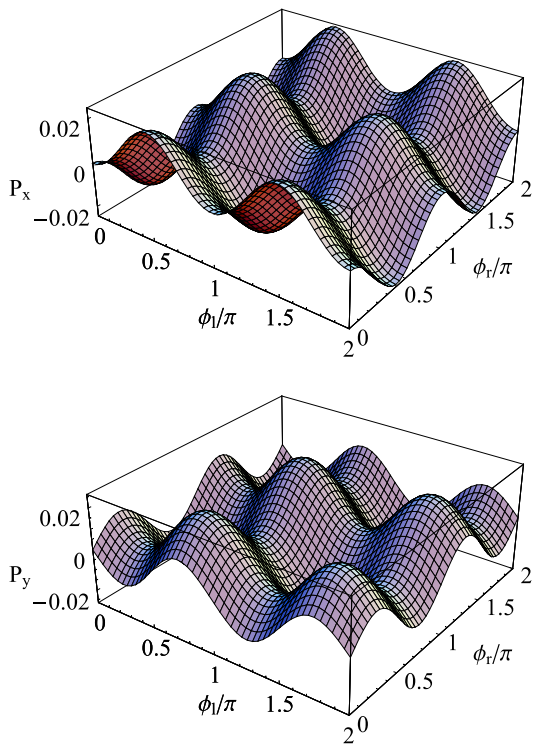}
\end{center}
\caption{(color online) Induced polarization $P_x$ and $P_y$ in a
unit of $L$ for the superexchange case with two holes in the
three-atom cluster. The parameters used in this calculation are the
same as in Fig. \ref{polarization}. } \label{superexchange-P}
\end{figure}

When two holes are introduced in the cluster, the ground state can
be changed adiabatically  without a level crossing as the two spin
directions are varied, and hence the polarization
exhibits a smooth change. Typical numerical results for the
longitudinal and transverse components of the polarization are shown
in the upper and lower panels of Fig.~\ref{superexchange-P}. As
in the previous consideration for one hole, the spin directions
are varied within the XY plane with
$\hat{m}_l=(\cos\phi_l,\sin\phi_l, 0)$ and
$\hat{m}_r=(\cos\phi_r,\sin\phi_r,0)$.

We analyzed the directional dependence of the polarization by using the
following forms:

\ba P_x /L &=& A \left[ \cos (2\phi_r ) - \cos (2\phi_l ) \right]
\nn
P_y /L &=& -B_1 \sin (\phi_r \!-\! \phi_l ) + B_2 \left[ \sin
(2\phi_r ) - \sin (2\phi_l )\right] . \nn \label{empirical-F}\ea
While the least-square fit to the numerical results were made using
as many as six different basis functions, only those listed above
made a sizable contribution. The angle dependence of the transverse
polarization $P_y$ predicted in Ref. \onlinecite{KNB} corresponds to
the $B_1 \sin (\phi_r - \phi_l )$ term above, whereas the angle
dependence of $P_x$ is consistent with the single-hole results
analytically derived in the previous section. The second term in
$P_y$, proportional to $B_2$, was not anticipated in analytic
theories but show up in the numerical analysis. In the spiral phase
for the spins, $B_1 \sin (\phi_r -\phi_l)$ produces a uniform
transverse component while all other terms give rise to a
macroscopically vanishing polarization. In this regard the new terms
($A$ and $B_2$) correspond to ``oscillating" component of the
polarization in contrast to the uniform component given by $B_1$.

%\begin{widetext}
\begin{figure}[t!]
%\begin{center}
%\includegraphics[width=9cm,angle=-90]{fourier.eps}
\includegraphics[width=6cm]{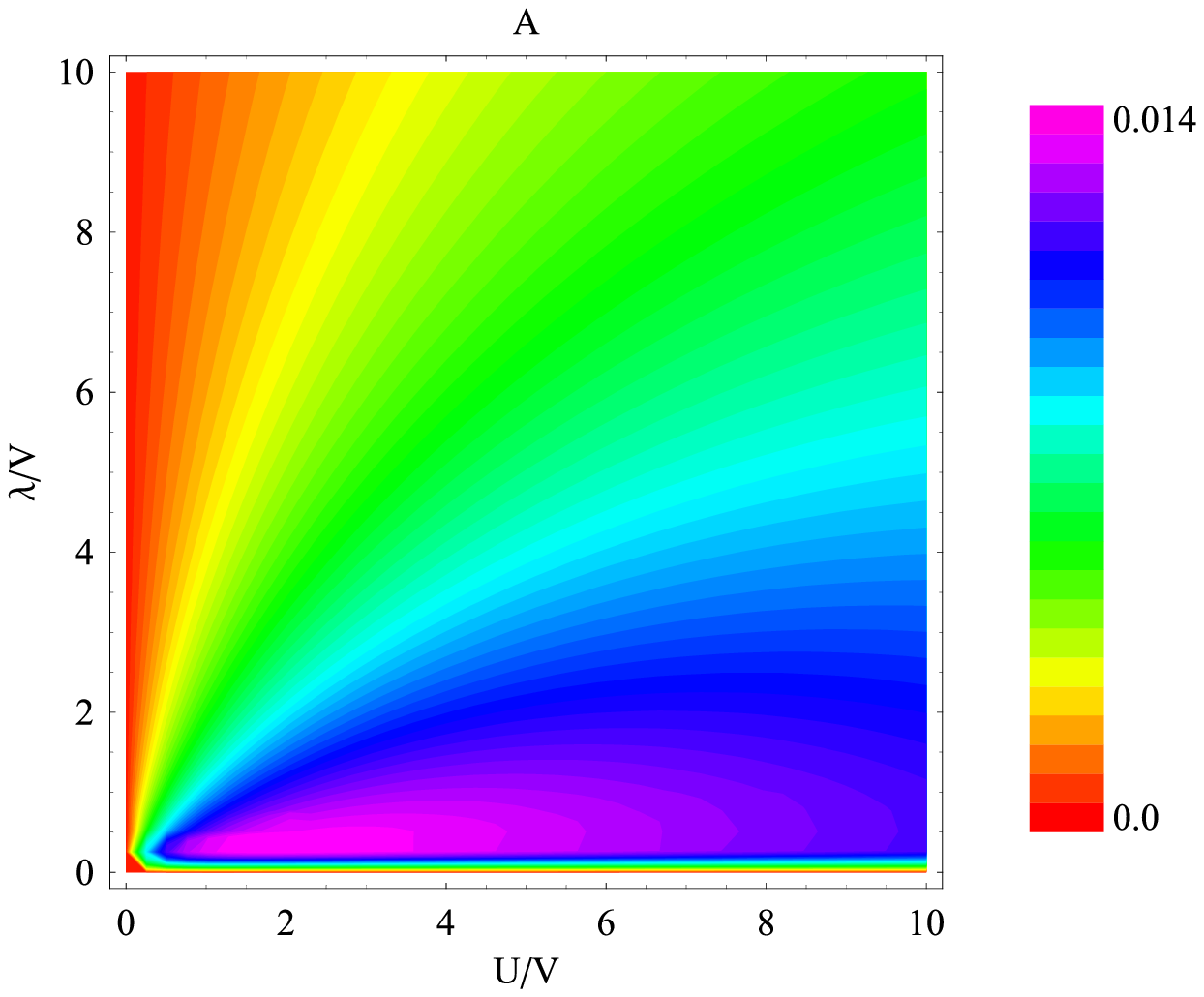} \\[10pt]
\includegraphics[width=6cm]{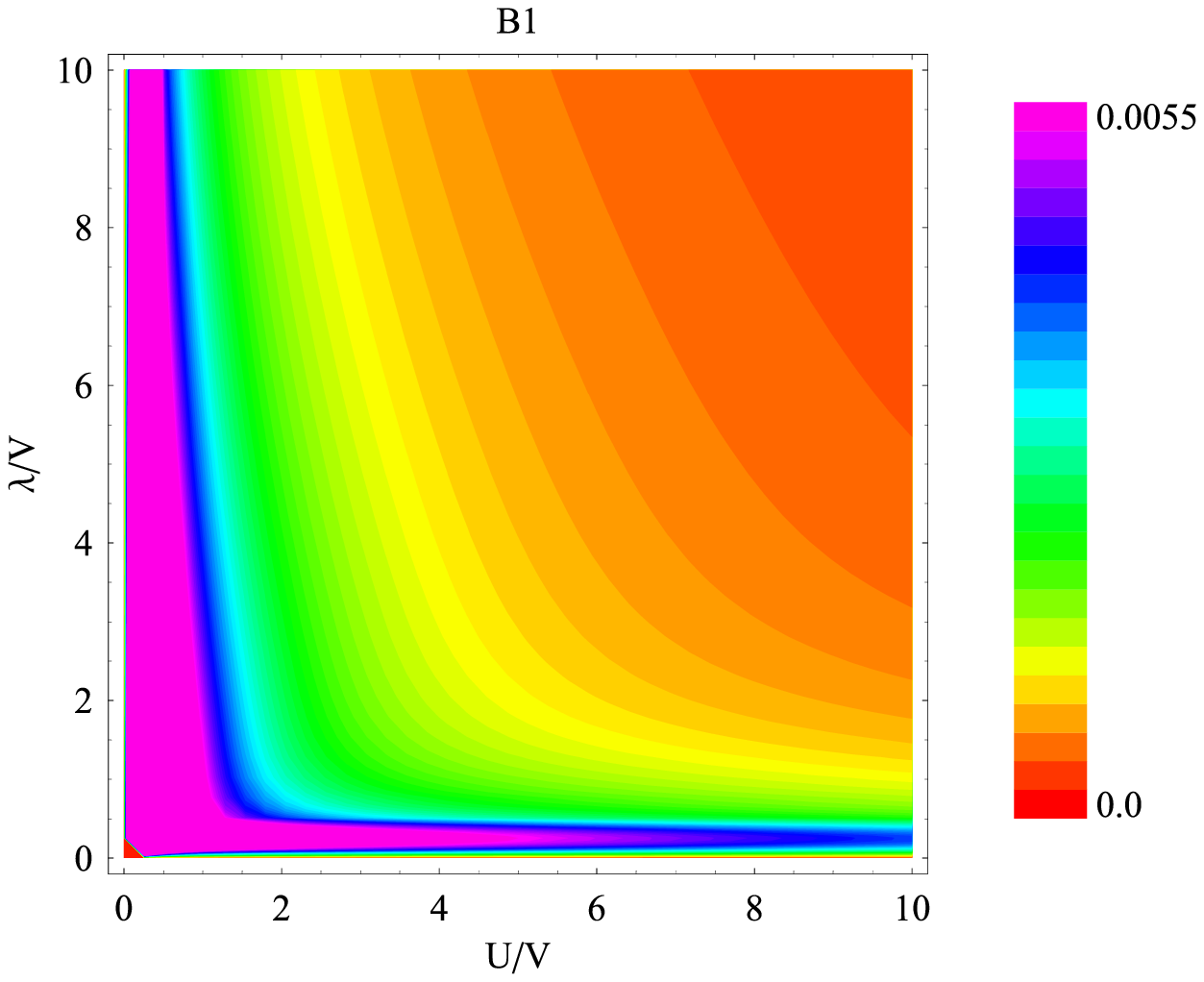}\\[10pt]
\includegraphics[width=6cm]{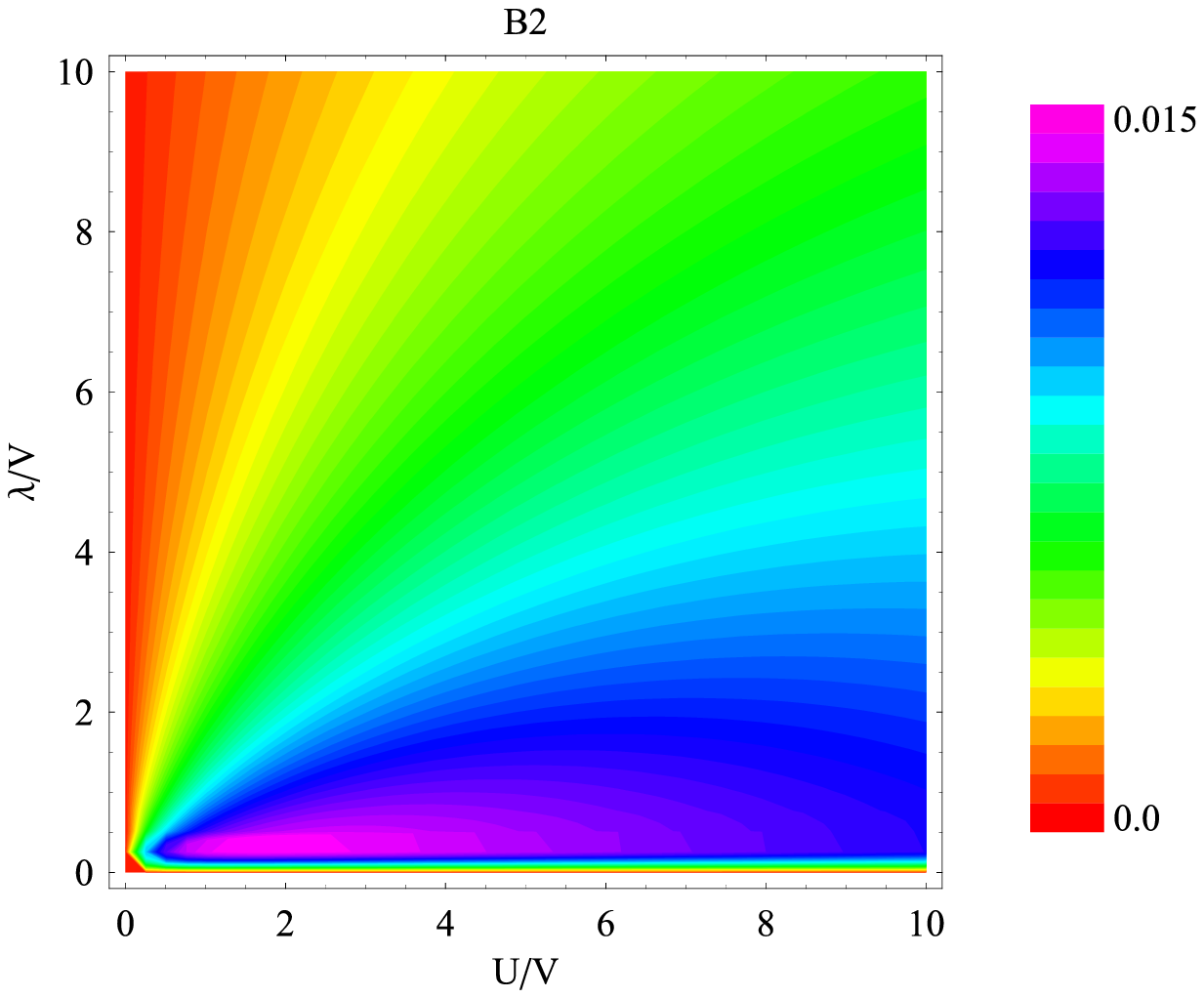} \\[10pt]
\includegraphics[width=7cm]{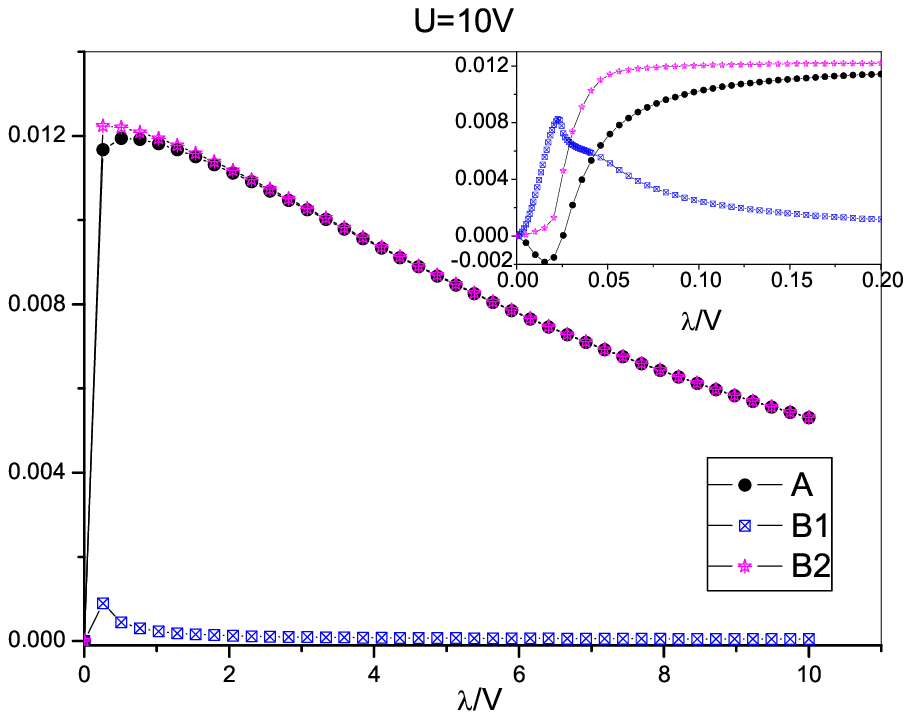}
\caption{(color online) (top three figures) Fourier components $A$,
$B_1$, and $B_2$, in the polarization $P_x /L$ and $P_y /L$ using
the fitting formula, Eq. (\ref{empirical-F}). (bottom) Fourier
components with $U/V=10$ and varying $\lambda/V$.} \label{fourier}
\end{figure}
%\end{widetext}

Shown in Fig. \ref{fourier} are the coefficients, $A$, $B_1$, and
$B_2$, for a wide range of parameters, $0 \le \lambda/V \le 10$, $0
\le U/V \le 10$, obtained by fitting the numerical data using Eq.
(\ref{empirical-F}).  Several features are worth mentioning: (i) $A$
and $B_2$ are remarkably similar over most of the parameter space.
This implies that the non-uniform polarization develops with similar
magnitudes along and orthogonal to the bond axis. (ii) Large
non-uniform polarization components are attained for large $U/V$ and
small $\lambda/V$ as indicated by the blue and violet regions in
Fig. \ref{fourier}. On the other hand, large uniform polarization
($B_1$) is achieved for large $\lambda/V$ and small $U/V$ in
agreement with Ref. \onlinecite{KNB}, where $\lambda \gtrsim U$ was
implicitly assumed. To be concrete, the largest values for $A$,
$B_1$ and $B_2$ are obtained for $(U/V, \lambda/V) = (1.795,0.256)$,
$(0.256,0.256)$, and $(1.794,0.256)$ respectively, with the
magnitudes indicated on the vertical bar next to each figure. Using
an earlier estimate\cite{KNB}, these maximal values correspond to
$A\sim B_2 \sim 2\times 10^2 ~\mathrm{n C/cm}^2$ and $B_1 \sim
80~\mathrm{n C/cm}^2$. For $U/V$ large, the polarization mainly
possesses the oscillating components with a much smaller uniform
component as explicitly demonstrated at the bottom of Fig.
\ref{fourier} for $U/V$ fixed at a large value ($U/V=10$) and
varying $\lambda/V$.

\begin{figure}[h]
\begin{center}
\includegraphics[width=8cm]{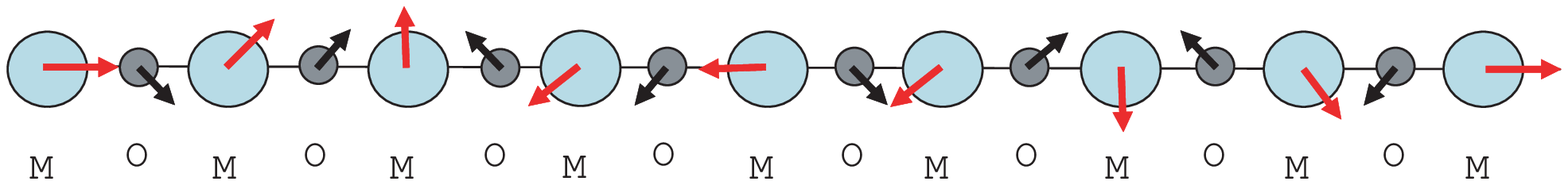}
\end{center}
\caption{Spiral spin structure (red arrows) with an angle shift of
$\pi/4$ between neighboring magnetic (M) ions and the induced dipole
moments (parallel or anti-parallel to black arrows) at oxygen (O)
sites with half the periodicity. Uniform component is not shown
since it is much smaller than the nonuniform part.} \label{o-shift}
\end{figure}

The macroscopic polarization is governed by the $B_1$ term alone. On
the other hand, the Fourier analysis reveals the local magnitudes of
the oscillating component to be much larger than the uniform
polarization over most of the parameter ranges, increasing the
chance to observe coupling of the local dipole order to external
probes such as X-ray or neutron scattering. Particularly in spiral
magnets with ordering wavevector $\bm{Q}$, the oscillation of
electric dipole moments appears with the wavevector $2\bm{Q}$, since
we have found that the nonuniform part has the form
$\sin2\phi_l-\sin2\phi_r$ for the transverse component and
$\cos2\phi_l-\cos2\phi_r$ for the longitudinal one.
Figure~\ref{o-shift} shows the simplest example with the periodicity
of the eight M ions where relative angles between the neighboring
spins are $\pi/4$. In this case, electric dipole moments induced at
oxyen ions also helically rotate with half the magnetic periodicity. Note that
the amplitude of this nonuniform part is much larger than the
uniform part. It is expected that oxygen ions subject to such
electric dipole field shit in either parallel or anti-parallel
directions to the local dipole moment.

\section{Discussion}
\label{discussion}

In this paper we have developed a theory for the spin-induced
electric polarization in a simple three-atom cluster model proposed
in Ref. \onlinecite{KNB}. The truncation to the low-energy space is
allowed by the large energy separation between spin-up and spin-down
states of the magnetic site due to the strong effective Zeeman
field, and the much smaller energy scale for the $pd$-hybridization
$V$ and spin-orbit interaction $\lambda$. The net electric dipole
moment is found to be along the cluster's axis within our
approximation, and the numerical calculation confirmed this
conclusion. The longitudinal polarization has an oscillating
character, and vanishes in the macroscopic limit. Numerical analysis
further identified uniform and non-uniform components in the
transverse polarization. The magnitude of the non-uniform transverse
polarization is comparable to the non-uniform longitudinal one. Such
non-uniform polarizations generically have larger magnitudes on the
atomic scale than the uniform transverse component.

In this paper we have concentrated on the $t_{2g}$ electron
systems because the spin-orbit interaction has the matrix
elements within the $t_{2g}$-orbitals, while they are zero
among the $e_{g}$-orbitals. The effective Zeeman field at
each site corresponds to the mean field from the
surrounding spins, which leads to the similar model to the
double exchange model discussed in this paper.
Generalizations to the more realistic models including the
hybridization between $e_{g}$- and $t_{2g}$-orbitals and
spin-orbit interaction on oxygen orbitals are now in
progress, which will be applied to the realistic materials
such as $R$MnO$_3$ ($R$=Tb,Gd). However, our present model
will have relevance to the materials such as YVO$_3$
\cite{Ren}, $R_2$Ti$_2$O$_7$ ($R$=Gd,Tb) \cite{py} which
have partially filled $t_{2g}$ electrons. More serious
considerations on each material including the finite bond
angle etc. are left for future investigations.

Recent experimental works on multiferroics with spiral spin
structure with the ordering wavevector $\bm{Q}$ have
reported the lattice modulation at the wavevector $2\bm{Q}$
in addition to the macroscopic polarization\cite{TbMnO3-2}.
Since dipole moments are linearly coupled to the ionic
displacements, it may correspond to the non-uniform dipole
moments shown in this paper. In our scenario of
magnetically induced dipole moments, the modulation of the
oxygen ions is non-collinear and has also a spiral
structure, reminiscent of the MnSi case~\cite{plummer}. A
similar lattice modulation but with the collinear shifts
has been discussed by Sergienko and Dagotto~\cite{dagotto}.
Further X-ray and neutron scattering experiments are
required to specify the detailed character of this lattice
modulation.

On the theory side, a Ginzburg-Landau consideration led to the
conclusion that the uniform polarization direction is orthogonal to
the modulation wavevector of the spin\cite{mostovoy,harris}. In Ref.
\onlinecite{mostovoy} some non-uniform components of the
polarization is predicted on the basis of Ginzburg-Landau theory
without a detailed discussion of its implications. Furthermore,
consideration of the orbital structure for the magnetic site is
missing in the phenomenological theories advanced so far. For
instance, the t$_{2g}$ levels alone break the rotational symmetry in
the orbital space and the corresponding Ginzburg-Landau theory need
not possess the invariance under spatial rotation. It will be
interesting to develop a Ginzburg-Landau theory with a proper
symmetry consideration for the $t_{2g}$ systems to see if the new
features we observe, i.e. non-uniform longitudinal and transverse
polarizations, emerge naturally.

Our results may also open up a possibility for a nanoscale
device of switching the polarization by means of an applied
magnetic field or the magnetization by means of an applied
electric field. The so-called ``quantum magnet'' might be a
candidate to such systems. The present results even pave a
way to fabricate crystaline magnets, ferroelectrics and
multiferroics having the longitudinal polarization.

\acknowledgments  The authors thank T. Arima for fruitful
discussions. One of us (C. Jia) was supported by the
Postdoctoral Research Program of Sungkyunkwan University
(2006). The work was partly supported by Grant-in-Aids
(Grant No. 15104006, No. 16076205, and No. 17105002) and
NAREGI Nanoscience Project from the Ministry of Education,
Culture, Sports, Science, and Technology.

\appendix*
\section{Calculation of the dipole moments}

In this appendix, we give details of analytic calculations
of dipole moments associated with the approximate
eigenstates of the total Hamiltonian.

\subsection{Diagonal terms}

Since

\begin{equation}
\langle Y_{1}|\v r|D_{1,xy}\rangle = \frac{L}{2\sqrt{1+|\kappa |}}\left[
1-1+|\kappa |-|\kappa |\right]\hat{x} = 0 ,
\end{equation}
we have

\ba && \langle \psi _{1}|\v r|\psi _{1}\rangle = -\frac{\sin \beta
_{1}}{2}\left[
\langle D_{1,xy}|\v r |Y_{1}\rangle +\langle Y_{1}|\v r|D_{1,xy}\rangle %
\right] =0,  \nn
&& \langle \varphi_{1}|\v r|\varphi _{1}\rangle = \frac{\sin \beta_{1}}{2}%
\left[ \langle D_{1,xy}|\v r |Y_{1}\rangle +\langle Y_{1}|\v
r|D_{1,xy}\rangle \right] =0 .  \nn \ea
All other diagonal terms vanish as well, $\langle \psi _{m}|\v r
|\psi _{m}\rangle =\langle \psi _{m}^{\prime }|\v r|\psi
_{m}^{\prime }\rangle =\langle \varphi _{m}|\v r|\varphi _{m}\rangle
=\langle \varphi _{m}^{\prime }|\v r|\varphi _{m}^{\prime }\rangle
=0$.
\\

\subsection{Off-diagonal terms}

In view of the symmetry of the
orbital overlaps given in Eq. (\ref{non-zero-overlap}), the
following averages must be zero: $\langle \psi _{1}| \v r|\psi
_{1}^{\prime }\rangle =\langle \psi _{1}|\v r|\psi_{2}^{\prime
}\rangle =\langle \psi_{2}|\v r|\psi_{1}^{\prime }\rangle =\langle
\psi_{2}|\v r|\psi_{2}^{\prime }\rangle =0$.
\\

\subsection{Dipole moments for the eigenstates}

Thus, only $\langle
\psi_{1}|\v r|\psi _{2}\rangle $, $\langle \psi^{\prime }_{1}|\v
r|\psi^{\prime }_{2}\rangle $ and $\langle \varphi _{1}| \v
r|\varphi _{2}\rangle $, $\langle \varphi^{\prime }_{1}| \v
r|\varphi^{\prime }_{2}\rangle $ may contribute to the polarization.
Since

\ba &&\langle D_{1,xy}|\v r|Y_{2}\rangle =-\hat{x}\times
L\sqrt{1-|\kappa |} \nn &&\langle Y_{1}|\v r|D_{2,xy}\rangle
=-\hat{x}\times L\sqrt{1+|\kappa |}, \ea
we have

\ba &&\langle \psi _{1}|\v r|\psi _{2}\rangle =  \nn
&&-\cos {\frac{\beta _{1}}{2}}\sin \frac{\beta _{2}}{2}\langle D_{1,xy}|%
\v r|Y_{2}\rangle -\sin \frac{\beta _{1}}{2}\cos {\frac{\beta _{2}}{2}}%
\langle Y_{1}|\v r|D_{2,xy}\rangle   \nn && \simeq \hat{x}\times
\frac{LV}{\Delta }\left[ \left( 1-|\kappa |\right) +\left( 1+|\kappa
|\right) \right] =\hat{x}\times \frac{2LV}{\Delta }, \nn \nn
&&\langle \varphi _{1}|\v r|\varphi _{2}\rangle =  \nn
&&\sin {\frac{\beta _{1}}{2}}\cos \frac{\beta _{2}}{2}\langle D_{1,xy}|%
\v r|Y_{2}\rangle +\cos \frac{\beta _{1}}{2}\sin {\frac{\beta _{2}%
}{2}}\langle Y_{1}|\v r|D_{2,xy}\rangle   \nn
&&\simeq -\hat{x}\times \frac{2LV}{\Delta }\sqrt{1-|\kappa |^{2}}=-\hat{x}%
\times \frac{\sqrt{2}LV}{\Delta }\sqrt{1-\sigma _{1}\cdot \sigma _{2}}.
\nonumber \\
&&
\end{eqnarray}
Similarly,

\ba &&\langle D_{1,zx}|\v r|Z_{2}\rangle =-\hat{x}\times
L\sqrt{1-|\kappa |} \nn
&&\langle Z_{1}|\v r|D_{2,zx}\rangle =-\hat{x}\times
L\sqrt{1+|\kappa |} \ea
gives

\ba &&\langle \psi _{1}^{\prime }|\v r|\psi _{2}^{\prime }\rangle =
\nn
&&-\cos {\frac{\beta _{1}}{2}}\sin \frac{\beta _{2}}{2}\langle D_{1,zx}|%
\v r|Z_{2}\rangle -\sin \frac{\beta _{1}}{2}\cos {\frac{\beta _{2}}{2}}%
\langle Z_{1}|\v r|D_{2,zx}\rangle   \nn &&\simeq \hat{x}\times
\frac{2LV}{\Delta },  \nn
&&  \nn &&\langle \varphi _{1}^{\prime
}|\v r|\varphi _{2}^{\prime }\rangle = \nn
&&\sin {\frac{\beta _{1}}{2}}\cos \frac{\beta _{2}}{2}\langle D_{1,zx}|%
\v r|Z_{2}\rangle +\cos {\frac{\beta _{1}}{2}}\sin \frac{\beta _{2}}{2}%
\langle Z_{1}|\v r|D_{2,zx}\rangle   \nn &&\simeq -\hat{x}\times
\frac{\sqrt{2}LV}{\Delta }\sqrt{1-\sigma _{1}\cdot \sigma _{2}}. \ea

\end{document}